# Alkoxide-intercalated NiFe-layered double hydroxides magnetic nanosheets as efficient water oxidation electrocatalysts

Jose A. Carrasco,[a] Jorge Romero,[a] María Varela,[b] Frank Hauke,[c] Gonzalo Abellán[a,c,*], Andreas Hirsch[c] and Eugenio Coronado[a,*]

[a] Instituto de Ciencia Molecular (ICMol), Universidad de Valencia, Catedrático José Beltrán 2, 46980, Paterna, Valencia, Spain.
[b] Universidad Complutense de Madrid, Dpt. Fisica Aplicada III & Instituto Pluridisciplinar, Madrid 28040, Spain.
[c] Department of Chemistry and Pharmacy and Institute of Advanced Materials and Processes (ZMP), University Erlangen-Nürnberg, Henkestr. 42, 91054 Erlangen and Dr.-Mack Str. 81, 90762 Fürth, Germany.

## Abstract

Alkoxide-intercalated NiFe-layered double hydroxides were synthesized *via* the nonaqueous methanolic route. These nanoplatelets exhibit high crystalline quality as demonstrated by atomic resolution scanning transmission electron microscopy combined with electron energy-loss spectroscopy. Moreover, the presence of the alkoxide moieties has been unambiguously demonstrated by means of thermogravimetric analysis coupled to a mass spectrometer. These NiFe–LDHs can be exfoliated in water or organic solvents and processed into homogeneous ultra-thin films (< 3nm thick) with the assistance of $O_2$-plasma. The study of their behaviour as water oxidation electrocatalysts has shown an outstanding performance at basic pHs (small overpotential of *ca.* 249 mV and Tafel slopes in the range of 52–55 mV per decade).

## Introduction

Since the discovery of graphene,[1] two dimensional (2D) materials have attracted a widespread attention from the scientific community,[2] including layered chalcogenides[3], boron nitride[4] or black phosphorus.[5] In addition to these van der Waals layered solids, other compounds formed by ionic layers can also provide examples of these 2D materials. An extensive family of this type is provided by the layered double hydroxides (LDHs), a class of anionic clays known the since the mid-19th century[6] which have been of interest in catalysis, sensing or magnetism.[7–10]

LDH can be formulated as $[M^{II}_{1-x} M^{III}_{x}(OH)_2]^{x+} (A^{n-})_{x/n} \cdot mH_2O$, in which $M^{II}$ and $M^{III}$ are divalent and trivalent metals, respectively (like $M^{II}$ = $Mg^{2+}$, $Zn^{2+}$, $Ni^{2+}$, $Co^{2+}$, and $Fe^{2+}$ or $M^{III}$ = $Al^{3+}$, $Fe^{3+}$, $Co^{3+}$, $Ni^{3+}$ or $Cr^{3+}$), and $A^{n-}$ is the interlayer anion which can be organic or inorganic and is placed between the cationic hydroxide sheets in order to compensate the excess of positive charge. One of the main features of these systems is their high chemical tunability, which allows them to be synthesized with different metallic compositions without altering their structure, leading to a wide range of properties. Moreover, the facile exfoliation of these LDH into monolayer nanosheets has boosted the attention paid to these materials.[11]

Among all the different possible compositions they exhibit, the combination of Ni and Fe (NiFe–LDHs) seems to be the most promising on the basis of its recent successful application as a highly efficient carbon monoxide (CO) oxidation catalyst, [12] their excellent behaviour as electroactive material in the water photolysis,[13–15] their use on high performance batteries[16,17] or as catalytic precursors for the CVD synthesis of novel carbon nanoforms.[18–21]

The synthesis and exfoliation of pure NiFe–LDHs with high crystallinity and well-defined hexagonal shapes have remained elusive so far.[22,23] Currently, there are two main routes for the synthesis of pure NiFe–LDHs that can be exfoliated into 2D nanosheets, namely the hydrothermal approach using urea as ammonium releasing reagent and triethanolamine as chelating reagent,[23,24] and the use of anthraquinone-2-sulfate that favors the topochemical oxidation of the metals during the layer formation.[25,26] These aqueous methodologies avoid the Fe oxidation, leading to NiFe–LDHs without spinel impurities.[27] When it comes to non-aqueous routes Gardner *et al.* described a synthetic procedure to obtain nanometric alkoxide-intercalated Al-containing LDHs using alcohols as solvents.[28,29] In this sense, we recently applied this method to synthesize pure CoFe–LDHs that can be exfoliated in water and exhibit excellent electrochemical properties.[30] In the present work, we have extended this non-aqueous route for the synthesis of NiFe–LDHs, obtaining alkoxide-intercalated nanoplatelets that exhibit size-dependent magnetic properties. Furthermore, we have explored their exfoliation in water and developed their processing into ultra-thin films with homogeneous coverage. Finally, we have characterized their electrochemical behavior as OER electrocatalysts showing an outstanding performance in alkaline solution.

**Experimental**

**Chemicals**

$NiCl_2 \cdot 6H_2O$, $FeCl_3 \cdot 6H_2O$, $Ni(NO_3)_2 \cdot 6H_2O$, $Fe(NO_3)_3 \cdot 9H_2O$, $C_6H_{15}NO_3$, $CO(NH_2)_2$, NaOH 97%, 1-Butanol, and Poly(vinylidene fluoride) (PVDF) were purchased from Sigma-Aldrich. Ethanol absolute, methanol (99.9%), and potassium hydroxide KOH (99.99%) were purchased from Panreac. Carbon black, acetylene 50% compressed, was obtained from Alfa Aesar (99.9%) and Iridium (IV) oxide from Stream chemicals. All chemicals were used as received. Ultrapure water was obtained from Millipore Milli-Q equipment.

**Synthesis of NiFe alkoxide (NiFe-A).**

The synthesis of the main sample was carried out following the method described by Gardner *et al*.[28,29] In a typical procedure, Ni and Fe salts were mixed in 100 mL of solvent (MeOH) in a molar ratio of 2:1 for a total of 40 mmol of metal cations. This solution was stirred and heated up to 65 °C for 1 h under Ar in a round bottom flask equipped with a reflux condenser. Afterwards, 3.8 g of NaOH dissolved in 100 mL of the MeOH was added dropwise to the initial mixture over a 2–3 min time span. The final mixture was left during 72 h at 65 °C and under magnetic stirring. Finally, the solution was filtered, washed thoroughly with MeOH and dried in vacuum. The final sample was labelled as NiFe-A.

[$Ni_{0.66}Fe_{0.33}(OH)_{1.55}(OMe)_{0.45}](Cl)_{0.33} \cdot$ 1 $H_2O$: (C, H, N, calc.: 4.2, 2.8, 0; found: 4.2, 3.1, 0.1).

**Synthesis of NiFe by hydrothermal approach (NiFe-HT).**

In a typical procedure, the nitrate salts were dissolved in 50 mL of Milli-Q water together with TEA, reaching a total metal cation concentration of 20 mM, with a ratio Ni:Fe of 2:1. TEA concentration was equimolar with Fe one. After that, 50 mL of an aqueous solution of urea (35 mM) was added. The final mixture was placed in a 125 mL stainless steel Teflon lined autoclave and heated up to 125 °C in an oven for 48 h. Afterwards, the autoclave was cooled to room temperature and the powder was filtered, washed with Milli-Q water and ethanol and dried in vacuum. The final sample was labelled as NiFe-HT.

[$Ni_{0.69}Fe_{0.31}(OH)_2](CO_3)_{0.155} \cdot$ 0.5 $H_2O$: (C, H, N, calc.: 2.0, 3.2, 0; found: 2.0, 3.6, 0.3)

**Exfoliation and deposition on $SiO_2$-Si substrates**

Tipically, $SiO_2$-Si substrates were cleaned and treated with oxygen plasma.[31]

For the deposition of exfoliated platelets in water, 1 mg of the sample was dissolved in 10 mL of Milli-Q water and sonicated for 30 min. Then, two approaches were considered, drop casting and spin coating (5000 rpm) of the solution onto the previously cleaned and treated with $O_2$ plasma $SiO_2$-Si wafer. On the other hand, for a complete coverage of the substrate a different approach was followed.[31] Typically, 10 mg of the NiFe-A LDHs were dissolved in a mixture of 20 mL of 1-BuOH and 10 mL of formamide in a covered flask. The solution was placed in an ultrasonic bath for 30 min. Then, the $SiO_2$-Si substrate was submerged into the solution for 5 min, followed by 1 min in a solution of 1-BuOH to remove the excess of LDH platelets. Afterwards, the substrates were dried at 70 °C overnight and stored in vacuum.

**Instrumentation**

X-ray powder diffraction (XRPD) patterns were obtained with a Philips X'Pert diffractometer using the copper radiation (Cu-Kα = 1.54178 Å). Field emission scanning electron microscopy (FESEM) studies were carried out on a Hitachi S-4800 microscope at an accelerating voltage of 20 kV and 30 seconds of Au/Pd metallization of the samples. ATR Infrared spectra were recorded with an Agilent Cary 630 FTIR spectrometer in the 4000–650 $cm^{-1}$ range with no need of KBr pellets. Scanning transmission electron microscopy and electron energy loss spectroscopy (STEM/EELS) characterization of the samples were carried out with a JEOL ARM200cF at University Complutense of Madrid, Spain, equipped with an aberration corrector, a cold field emission gun, and a Gatan Quantum spectrometer. Samples were prepared by dropping a colloidal suspension of the fresh sample in EtOH on a holey carbon-coated copper grid for STEM-EELS observation. Thermogravimetric analysis (TGA) coupled with a mass spectrometer (MS) was performed on a Netzsch STA 409 CD instrument equipped with a Skimmer QMS 422 mass spectrometer (MS/EI) with the following programmed time-dependent temperature profile: 25–500 °C with 10 °C·$min^{-1}$ gradient and cooling to room temperature. The initial sample weights were about 5 mg, and the whole experiment was performed under helium with a gas flow of 80 mL·$min^{-1}$. Atomic force microscopy (AFM) measurements were performed with a Multimode atomic force microscope (Veeco Instruments, Inc.). The images were obtained with a Si tip (frequency and K of ≈300 kHz and 42 N·$m^{-1}$, respectively) using the tapping-mode in air at room temperature. Images were recorded with 512 X 512 pixel and a 0.5−1 Hz scan rate. Processing and analysis of the images were carried out using the Nanotec WSXM-5.0 Develop 6.0 software (www.nanotec.es).[32] Dynamic light scattering (DLS) measurements were recorded at 25 °C with a Zetasizer Nano ZS instrument from Malvern Instrument Ltd on the aqueous suspensions previously described. Magnetic data were collected with a Quantum Design superconducting quantum interference device (SQUID) MPMS-XL-5. The susceptibility data were corrected from the diamagnetic contributions of the atomic constituents of the samples as deduced from Pascal's constant tables and the sample holder. The dc data were obtained under an external applied field of 100 or 1000 Oe in the 2–300 K temperature range. Magnetization studies were performed between −5 and +5 T at constant temperatures of 2 and 20 K.

**Electrochemical measurements**

The electrochemical experiments were performed using an Autolab electrochemical workstation (PGSTAT-100 potentiostat/galvanostat) connected to a personal computer that uses GPES electrochemical software.

The powdered materials were mixed with acetylene black and PVDF in a mass ratio of 80:10:10 in ethanol and deposited on a nickel foam electrode. The as-prepared nickel foam electrodes were dried overnight at 80 °C and pressed. Each working electrode contained about 0.25–0.5 mg of electroactive material and had a geometric surface area of about 1 cm$^2$. The synthesis of the *in situ* grown films was performed by following the same experimental procedure described for NiFe-A but replacing the magnet used for the stirring with Ni foam leading to the direct growth of NiFe-A on this material.[33] A typical three-electrode experimental cell equipped with a stainless steel plate having 4 cm$^2$ of surface area as the counter electrode, and a Metrohm Ag/AgCl (3 M KCl) as the reference electrode was used for the electrochemical characterization of the working electrodes. All measurements were carried out with magnetic agitation and nitrogen bubbling.

The electrochemical properties were studied measuring the CV at different scan rates in 1 M KOH aqueous solutions. In addition, chronoamperometric studies were performed at a constant overpotential (j = 0.3 V), and chronopotentiometric studies at constant current densities of 5 and 10 mA·cm$^{-2}$. All potentials reported in this manuscript were converted to the RHE reference scale using E(RHE) = E(NHE) + 0.059·pH = E°(Ag/AgCl) + 0.197 V + 0.059·pH. The turnover frequency (TOF) values were calculated from the equation: $TOF = JA/4Fm$

Where *J* is the current density at a given overpotential of 0.3 V, *A* is the surface area of the working electrode, *F* is the Faraday constant and *m* is the number of moles of metal loaded on the electrode.

**Results and discussion**

The synthesis of NiFe-LDH alkoxide (NiFe-A) was carried out using a methanolic solution including the metal chloride salts under basic pH at 65 °C, following a modified method reported by Gardner and co-workers.[28–30] After the 72 h of synthesis, the resulting brown precipitate was filtered and washed thoroughly with MeOH, dried in a vacuum chamber and stored under vacuum in order to avoid carbonate contamination.

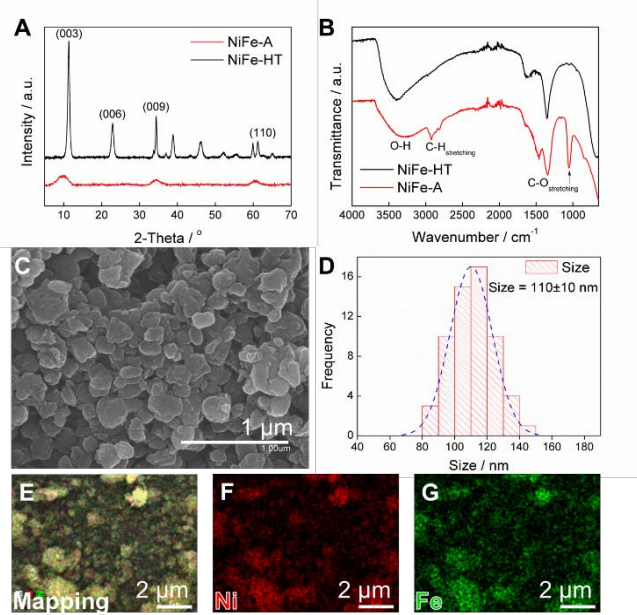

**Fig. 1** XRPD (A) and FTIR spectra (B) of the NiFe-A and NiFe-HT samples. FESEM image of NiFe-A (C). Histogram showing the average particle size distribution for the NiFe-A sample (D). Mapping analysis of the NiFe-A highlighting the homogeneous distribution of Ni and Fe in the layers (E, F, and G).

The synthesis of the LDH phase was confirmed by X-ray powder diffraction (XRPD) analysis (Figure 1A), showing the characteristic (110) doublet at around 60° in 2θ, and the main basal reflection (003) at around 2θ = 9–9.5°, providing a basal space of *ca.* 9.3 Å, in goods agreement with the expected values for MeO-intercalated LDH.[28] For the sake of comparison the XRPD spectrum of the NiFe-LDH obtained by hydrothermal approach (NiFe-HT) was also depicted (Figure 1A), showing the (003) reflection at a 11.4° indicative of the presence of carbonate anions in the interlamellar space.[20,23]

The Fourier transform infrared spectra (FTIR) of NiFe-A and NiFe-HT are compared in Figure 1B. There are two bands in the former case at *ca.* 2950 and 1070 cm$^{-1}$, related to C-H and C-O stretching vibrations, respectively., which are absent in the carbonate-intercalated sample, supporting the presence of the alkoxide anion in the interlamellar space.[34]

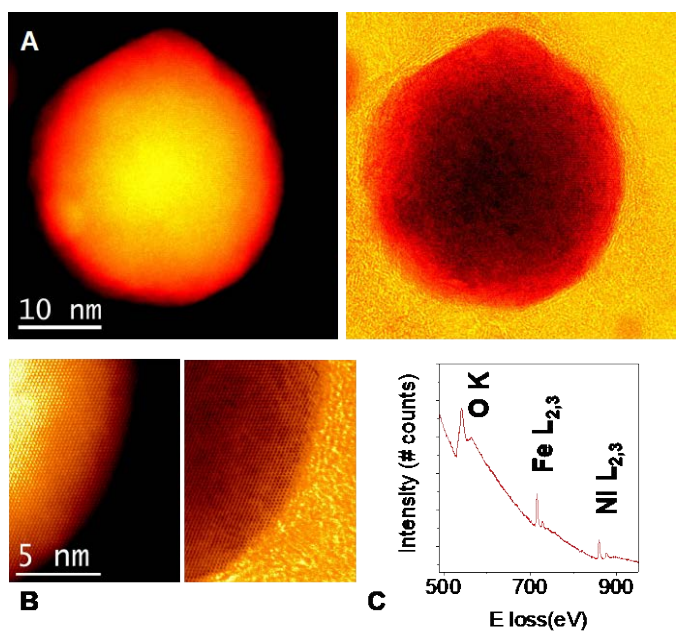

**Fig. 2** (A) Annular dark field (ADF) and simultaneously acquired annular bright field (ABF) low magnification images –left and right respectively– of a small NiFe-A flake. The flake is monocrystalline and the crystal quality is very high. (B) High magnification ADF (left) and ABF (right) images of the flake edge, exhibiting a high degree of crystallinity. No major defects are observed. (C) EEL spectra of a flake, showing the O K, Fe $L_{2,3}$ and the Ni $L_{2,3}$ absorption edges. STEM images acquired at an acceleration voltage of 200 kV, EEL spectrum measured at 80 kV on a different flake.

FESEM measurements were carried out to unveil the morphology of both samples (Figure 1C). NiFe-A depicts homogeneous lateral dimensions of around 110 nm, in clear contrast to NiFe-HT, which depicts micrometric sizes (See SI 1 and 2 for additional FESEM images).[24] It is worth to remark that, whereas the morphology of the NiFe-HT consists on well-defined hexagonal shapes, NiFe-A does not show this good definition, resulting into more irregular flakes.

Similar dimensions have been reported for the MgAl derivative by Gursky *et al.*[29], or conventional NiAl-HT.[35] For CoFe-A synthesized with our non-aqueous method, values of *ca.* 20 nm were reported.[30] EDAX mapping confirmed the expected 2:1 Ni:Fe ratio showing a homogeneous distribution of the two metals throughout the whole sample (Figure 1E–G). Furthermore, elemental analysis, microanalysis, and thermogravimetric analysis of the sample allowed us to estimate the following molecular formula for NiFe-A: $[Ni_{0,66}Fe_{0,33}(OH)_{1,55}(OMe)_{0,45}](Cl^-)_{0,33}\cdot 1\ H_2O$.

Direct information of the microstructure of the nanoplatelets can be obtained by aberration-corrected STEM-EELS. Fig. 2A shows a high angle annular dark field (HAADF) STEM image of a single NiFe-A flake. The platelet-like nanocrystal has a lateral dimension of ~35 nm, the smaller size is related with the rupture of the flake after the sonication used for the preparation of the TEM grid. Fig. 2B depicts a high magnification atomic resolution image showing a high quality crystalline structure. In order to confirm the chemical composition of the sample, the corresponding EELS spectrum was obtained working at 80 kV and illuminating a crystal while scanning the electron beam in order to minimize the beam-induced damage. The O K, Fe $L_{2,3}$ and Ni $L_{2,3}$ edges are visible, near 530 eV, 708 eV, and 855 eV, respectively. The quantification of the spectra using the routine available in the Gatan Digital Micrograph software (and hydrogenic-white line cross-sections) yields a Ni:Fe atomic ratio of 1:(0.61± 0.09), in good agreement with EDAX microanalysis.

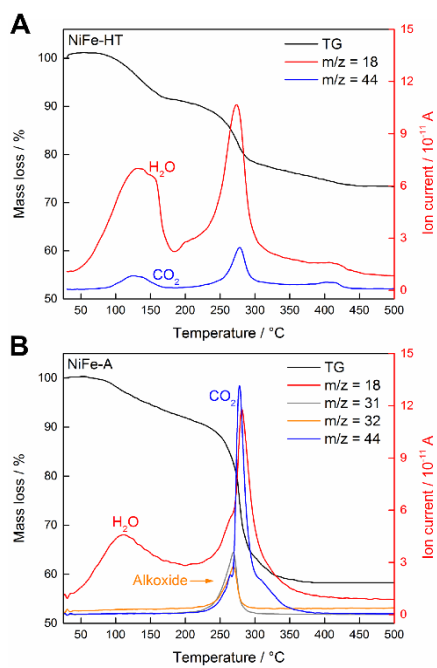

**Fig. 3** (A) TGA-MS analysis of NiFe-HT using a heating rate of 10 K·min$^{-1}$ under helium. The mass traces corresponding to H$_2$O (m/z 18) and CO$_2$ (m/z 44) are highlighted. The intense peak mass loss takes place at 278 °C. (B) TGA-MS analysis of NiFe-A showing, in addition to the H$_2$O and CO$_2$, the MS traces attributed to the intercalated methoxide fragments – m/z 31 and 32.

To provide further insight into the chemical nature of the intercalated species we have characterised, for the first time, the NiFe-LDH samples with thermogravimetric analysis coupled with a mass spectrometer (TG-MS) under an inert atmosphere of helium (Figure 3). This powerful technique allows us to unambiguously identify the interlamellar anions, confirming the intercalation of the alkoxide groups.[36] Analysis of NiFe-HT from room temperature to 500 °C revealed the presence of carbonate (CO$_2$, m/z 44) with a continuous mass loss peaking at *ca*. 280 °C. With respect to the H$_2$O (m/z 18), a first intense mass loss attributed to the weakly physisorbed water appeared between room temperature and *ca*. 180 °C, then a second and more intense mass loss takes place at around 274 °C and is correlated with the dehydroxylation of the layers as well as with the detachment of chemisorbed water present in the interlamellar space (Figure 3A).[36] In the case of NiFe-A the continuous MS enabled the allocation of the molecular fragments corresponding to the alkoxide moiety, m/z 31 and 32, at around 270 °C, a temperature remarkably higher than that of the pristine MeOH (*ca*. 65 °C). Interestingly, no signal of $CO_2$ was detected till the appearance of a tiny peak centered at *ca.* 264 °C, indicative for the residual character of the carbonate contamination in the alkoxide-intercalated samples as recently observed for other hybrid MgAl- and ZnAl-LDHs.[36] Its worth to remark that the $CO_2$ signal of NiFe-A at higher temperature is also related with the combustion of the organic anions (Figure 3B).

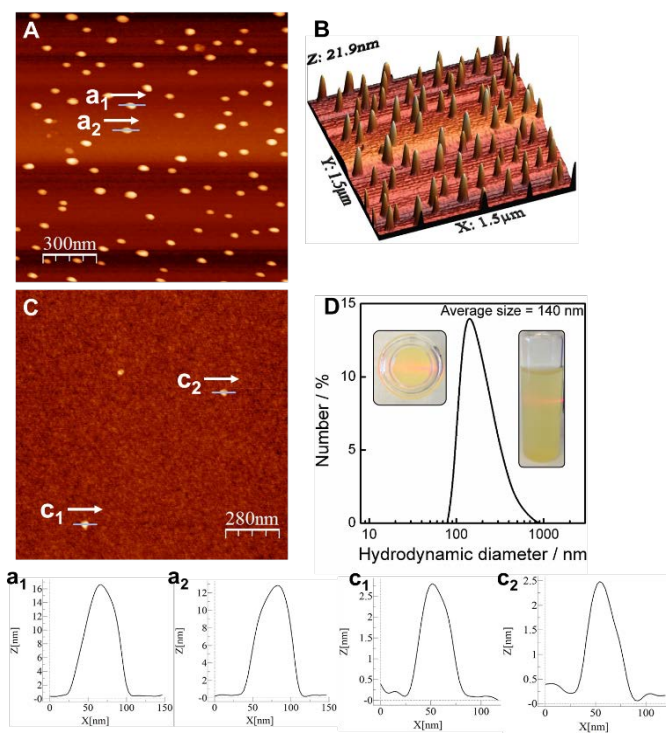

**Fig. 4** (A) AFM images of a solution of water exfoliated NiFe-A deposited by drop casting onto a $SiO_2$-Si substrate and (B) its 3D representation. (C) Topographic AFM image of a spin coating deposition of the previous solution. (D) DLS and Tyndall effect of the NiFe-A sample exfoliated in water. Profiles of the marked regions in A ($a_1$, $a_2$) and C ($c_1$, $c_2$).

One of the main features of LDH is their ability to be exfoliated in appropriate solvents.[11] In the case of the alkoxide-intercalated LDH is also possible to perform the exfoliation in water. Indeed, by suspending 1mg of LDH in 10 mL of Milli-Q water and sonicating during 30 min, a clear colloidal suspension showing Tyndall-Faraday effect can be prepared. Dynamic light scattering measurements provided an average hydrodynamic diameter of *ca.* 140 nm in excellent agreement with the FESEM studies (Figure 4). The exfoliation and further deposition of the NiFe-A sample on $SiO_2$-Si substrates was also confirmed *via* AFM measurements. We firstly performed a typical drop casting of the NiFe-A water suspension observing the accretion of the platelets into clusters of *ca.* 10–20 nm in thickness (Figure 4A–B). This situation can be overcome by spin coating the same solution at 5000 rpm (two times) lowering the thickness to *ca.* 2.5 nm (Figure 4C).

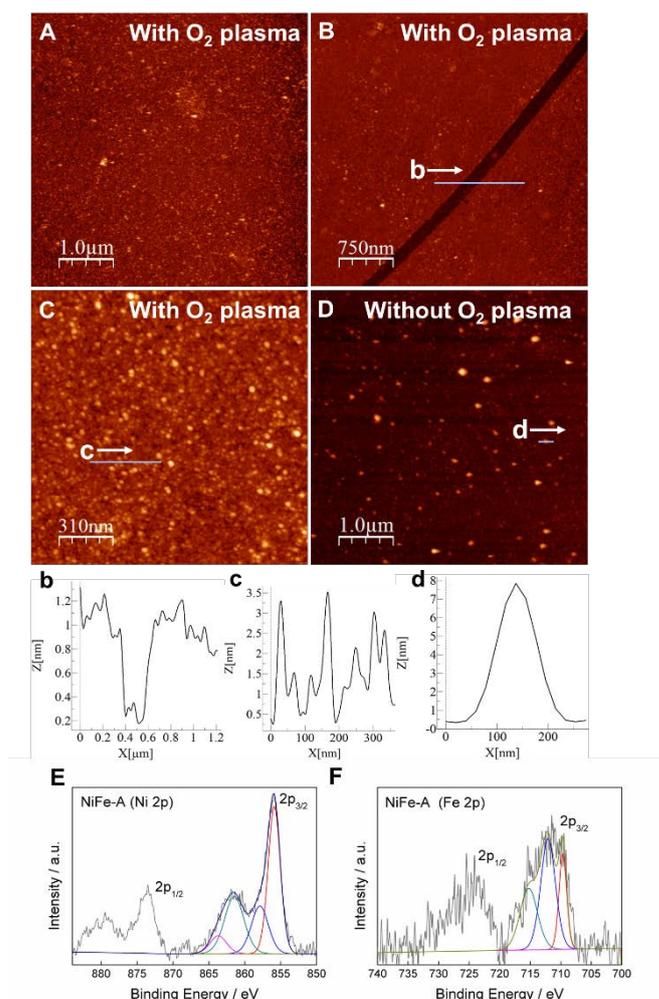

**Fig. 5** AFM images of the SiO$_2$-Si substrate after being covered with NiFe-A LDH dissolved in a mixture of formamide and butanol in the presence (A, B, C) or absence (D) of O$_2$ plasma activation of the film prior the deposition. Image (B) highlights a scratch in the surface revealing the overall thickness of the film. Profiles corresponding to the marked regions in B (b), C (c) and D (d). XPS analysis of the film highlighting the Ni 2p (E) and Fe 2p (F) high-resolution spectra.

The processability of LDH into thin films is one of the most rapidly growing areas in this field, as the large-scale correct disposition of the LDH platelets would facilitate their application in energy storage and conversion devices.[20,37] Along this front, we checked the formation of continuous thin films by physical deposition. The first attempts with the water suspensions were unsatisfactory due to an inhomogeneous coverage even after using O$_2$-plasma cleaning and activation of the substrates (Fig. 4). To face this challenge we used a solvent mixture of 1-buthanol and formamide, following a slight modification of the method reported by Jung et al.[31] including O$_2$-plasma activation. As seen in Figure 5, complete coverage of the substrate was achieved, with lateral sizes of ca. 50–100 nm for the platelets and small RMS values of 0.83 nm. The average height of the coverage is 2.67 nm, revealing a highly homogeneous film. The scratch of the film gave us an approximate thickness value, resulting in ca. 1 nm (average of the measurement in different points), very close to that expected for a monolayer. This film is remarkably thinner than that previously synthesized by in-situ growth of the NiFe-LDH on glass substrates (*ca.* 138–170 nm).[20] A control experiment on a cleaned substrate including its RMS value is depicted in the Supporting Information (SI 3). The role exerted by the oxygen plasma is crucial. Indeed, control experiments without using the $O_2$-plasma revealed poor deposition of the LDH, with higher thickness values that might be attributed to some agglomeration of the platelets (Fig. 5D). The plasma treatment allows us to activate the silicon substrate with negative charges,[38,39] leading to a better bonding with the positive LDH cationic sheets, and resulting in a homogeneous and large coverage (additional images can be seen in SI 4).

The chemical integrity of the resulting films was confirmed by XPS. Figure 5 E and F shows the high-resolution Ni 2p and Fe 2p XPS spectra of the film confirming the presence of Ni(II) and Fe (III) in the samples. The Ni 2p spectrum exhibit two main peaks at 855 and 880 eV, related to the spin-orbit splitting of the Ni ($2p_{3/2}$) and Ni ($2p_{1/2}$), respectively. In the case of Fe 2p, two peaks are centred at binding energies of 713 and 726 eV, indicative for Fe ($2p_{3/2}$) and Fe ($2p_{1/2}$), respectively. These binding energy values are in good agreement with those expected for NiFe-LDH phases.[20,40,14] Our new synthetic route represents a straightforward way for the preparation of homogeneous ultrathin films of NiFe-LDH.

The overall magnetism of a LDH system is controlled by two main contributions.[10] On the one hand the intralayer magnetic superexchange interactions between metallic centres through the hydroxyl (OH-) bridges across the cationic sheets. On the other hand, the less intense dipolar interactions, which take place in the space between the magnetic LDH layers (interlayer nature). In our specific case, NiFe-LDHs behave as low-temperature ferrimagnets due to the coexistence between ferromagnetic Ni-OH-Ni superexchange interactions along with antiferromagnetic Ni-OH-Fe and Fe-OH-Fe interactions.[20,24,30,27,41] All magnetic measurements were carried out in a SQUID with freshly prepared powdered samples. The main magnetic data and parameters have been summarized in Table 1, including NiFe-HT[24] and CoFe-A[30] for comparative purposes. The DC susceptibility measurements ($\chi_M$) depict a sharp increase near 50 K, reaching a maximum value of 0.63 emu·mol$^{-1}$ at 4.5 K, indicative of cooperative magnetic interactions (Fig 6A). On the other hand, the thermal variation of $\chi_M \cdot T$ decreases from a value of 1.93 emu·K·mol$^{-1}$ at room temperature to 1.86 emu·K·mol$^{-1}$ at 70 K. After that, $\chi_M \cdot T$ exhibits a sharp increase upon cooling to its maximum value of 4 emu·K·mol$^{-1}$ at 8.1 K, followed by an abrupt decrease to 1.23 emu·K·mol$^{-1}$ at 2 K (Fig 6A). Fitting the DC data to the Curie-Weiss law above 50 K gives rise to a Curie constant (C) of 1.95 emu·K·mol$^{-1}$ consistent with that expected for the spin only value of a magnetically diluted combination of $Ni^{2+}$ (S = 1) and $Fe^{3+}$ (S = 5/2) ions (Table 1).[24] Moreover, the positive value (4.74 K) of the Weiss constant (θ) is indicative for the predominance of ferromagnetic interactions throughout the layers, although its value is remarkably smaller than in the NiFe-HT. Field cooled and zero field cooled (FC/ZFC) measurements allowed us to extract both blocking and irreversible temperatures, resulting in $T_B$ = 4.8 K and $T_{irr}$ = 5 K, respectively (Figure 6B). Hysteresis loop were also recorded at 2 and 20 K –below and above the blocking temperature– (Fig 6C), confirming the presence of spontaneous magnetization at low temperatures. The coercive field increases as long as we decrease the temperature, as expected, with a $H_c$ of 690 Oe at 2 K and 30 Oe for 20 K. These $H_c$ values are always lower than 1000 Oe, concluding that this NiFe-A LDH is a soft magnet, in good agreement with what is found for the CoFe-A.[24,30] On the other hand, micrometric NiFe-HT depicted a higher coercive field of *ca.* 3600 Oe for the 2:1 ratio, as well as a much higher irreversible temperature (15.1 K) pointing towards superparamagnetic behaviour of the NiFe-A due to the dramatic decrease in the particle size.[24]

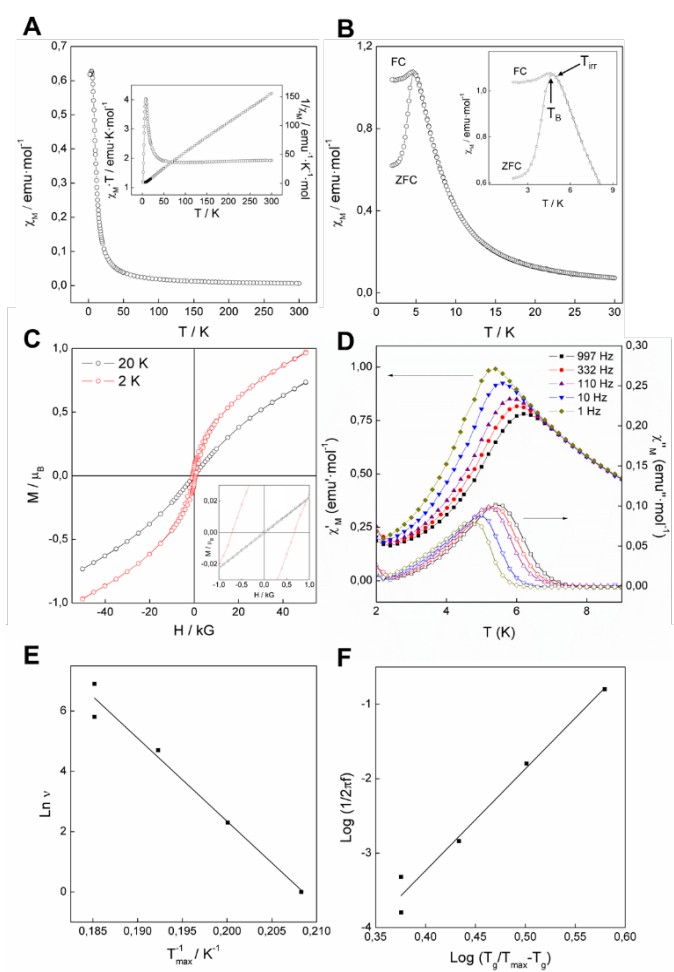

**Fig. 6** Magnetic properties of the NiFe-A sample. (A) $\chi_M$ *vs.* T plot. The inset in represents the temperature dependence of the $\chi_M \cdot T$ product and the fitting of the magnetic data to a Curie-Weiss law. (B) FC/ZFC plot, the inset remarks the low field region highlighting the irreversible (bifurcation point) and blocking temperature. (C) Hysteresis cycle at 2 and 20 K. The inset highlights the low field region. (D) Thermal dependence of the χ'$_M$ (in-phase) and χ"$_M$ (out-of-phase) signals at 1, 10, 110, 332 and 997 Hz.

(E) Arrhenius fitting of the χ"$_M$ signal. (F) Frequency dependence of χ"$_M$ fitted with the 3D scaling low model $\tau = \tau_0 \cdot \left[T_g/(T_g/T_{max} - T_g)\right]^{zv}$.

AC dynamic susceptibility measurements in the presence of an external field of 3.95 Oe oscillating at different frequencies in the 1 – 10000 Hz interval were carried out in order to confirm the cooperative magnetism in the sample (Fig. 6D). In both cases, the in-phase (χ'$_M$) and the out-of-phase (χ"$_M$) signals exhibit a defined peak at low temperatures.

From the out-of-phase signal we can extract the temperature for the onset of the spontaneous magnetization ($T_M$), set as the point where χ"$_M$ ≠ 0. For the NiFe-A $T_M$ = 7.8 K, remarkably lower than that observed for NiFe-HT ($T_M$ = 16.8 K), which is indicative of size effects.[24,42] As the experimental ratio is 2:1 between Ni:Fe (χ = 0.33), the $Fe^{3+}$ clustering cannot be avoided, as previously demonstrated by Mössbauer spectroscopy for NiFe-HT, leading to a glassy magnetic behaviour.[24] Along with that, both signals χ'$_M$ and χ"$_M$ exhibit a significant frequency dependence. The frequency dependence of the χ'$_M$ maxima can be estimated with the calculation of the frequency shift parameters defined by Mydosh[43,44]:

$$\Phi = \Delta T_{max}/[T_{max}\Delta(\log v)] \quad \text{Eq. (1)}$$

For NiFe-A, the value Φ ≈ 0.043 is higher than that observed for canonical spin glasses (0.005 – 0.018)[43] and within the range of values associated to spin glass-like materials (0.06 – 0.08). Notice that nanosized NiAl- , CoAl-[45] and CoFe-LDHs,[30] exhibit Φ values in this range, in line with the expected size effects. In contrast, in NiFe-HT Φ is in the range of values associated toclose to canonical spin glasses (Φ ≈ 0.021).[24] Further information of the spin relaxation in these materials can be obtained by fitting the frequency dependence of χ"$_M$ to both an Arrhenius law and a 3D critical scaling law.

The Arrhenius law for a thermally activated process is described by the following equation:

$$v = v_0 \cdot \exp(-E_a/k_B T) \quad \text{Eq. (2)}$$

where $v_0$ is the frequency factor, $E_a$ the activation energy and $k_B$ the Boltzmann constant.

For NiFe-A a fitting of the frequency dependence of χ"$_M$ to this law gives a value for energy barrier of $E_a/k_B = 276$ K, which can be ascribed to the presence of a superparamagnetic behaviour[43], in clear contrast to the NiFe-HT sample that exhibits a higher value of *ca.* 1400 K. These results highlight the crucial role of the nanometric size and low dimensionality of these LDH systems in sharp contrast with the micrometric ones.[46]

Finally, we have fitted the frequency dependence of χ"$_M$ to the 3D critical scaling law for spin dynamic,[47,48] which is described by the following equation:

$$\tau = \tau_0 \cdot [T_g/(T_g/T_{max} - T_g)]^{z\nu} \quad \text{Eq. (4)}$$

where $T_g$ is the critical glass temperature, $\tau_0$ the attempt time and $z\nu$ a critical exponent. For 3D model spin glasses, the relaxation time diverges at finite temperature ($T_g \neq 0$ K). For NiFe-A the best fit of our data to a linear form of the equation 4 (Fig 7F) was obtained for $T_g$=3.8 K, $\tau_0$= 1.99·10$^{-9}$ s and a $z\nu$=13.7. Whereas the obtained $\tau_0$ value falls in the range for canonical spin glasses (from 10$^{-7}$ to 10$^{-12}$), $z\nu$ is out of this range (from 4 to 12).[43]

A similar behaviour was reported by Layrac *et al.* for different magnetic LDHs intercalated with cyano-bridged coordination polymers.[47] For NiFe-HT, the best fit (see SI 5) was found for $T_g$=14.9 K, $\tau_0$= 2.55·10$^{-7}$ s and $z\nu$=2.7.[48] In overall, these results indicate a spin-glass like behaviour for these NiFe-LDHs, with superparamagnetic effects in the NiFe-A derivative arising from the nanometric size of the samples.[47,44]

To demonstrate the applicability of these novel NiFe-LDHs in the field of energy storage and conversion, we have investigated their performance as OER electrocatalysts for the multi-electron reaction: $4OH^- \leftrightarrow O_2 + 2H_2O + 4e^-$, in alkaline media.[49] Beyond other compositions, the catalysts containing earth-abundant Ni and Fe cations are among the most efficient reported so far.[50,51] Along this front, the performance exhibited by NiFe-LDH and its corresponding carbon hybrids have been studied.[14,15,52,26,53–55] a $\chi \cdot T_{rt}$ value at room temperature; expected spin-only value of the Curie constant ($C_{SO}$); experimental Curie constant (C); Weiss constant (θ); temperature of the divergence of the ZFC and FC magnetic susceptibility ($T_{irr}$); temperature for the onset of spontaneous magnetization extracted from the $\chi''_M$ plot ($T_M$); measured hysteresis temperature ($T_{Hys}$); saturation magnetization ($M_S$); coercive field ($H_{Coer}$); energy barrier ($\Delta/k_B$) and frequency factor ($\nu_0$), resulting from the fitting of the magnetic susceptibility to the Arrhenius law; Mydosh parameter ($\phi$). S(Ni$^{2+}$) = 1, S(Fe$^{3+}$) = 5/2.

**Table 1:** Main magnetic data and parameters for the magnetic LDH.[a]

| Sample | $\chi \cdot T_{rt}$ (emu·K·mol$^{-1}$) | $C_{SO}$ (emu·K·mol$^{-1}$) | $C$ (emu·K·mol$^{-1}$) | $\theta$ (K) | $T_{irr}$ (K) | $T_M$ (K) | $T_{Hys}$ (K) | $M_S$ ($\mu_B$) | $H_c$ (Oe) | $\Delta/K_B$ (K) | $\nu_0$ (Hz) | $\phi$ | Ref. |
|---|---|---|---|---|---|---|---|---|---|---|---|---|---|
| NiFe-A | 1.93 | 2.10 | 1.95 | 4.74 | 5.0 | 7.8 | 2 ; 20 | 0.97; 0.74 | 690; 30 | 275.7 | 9.1·10$^{24}$ | 0.043 | This work |
| CoFe-A | 2.48 | 2.50 | 2.57 | -14.69 | 4.8 | 7.0 | 2.0 | 0.98 | 402 | 116.9 | 3.8·10$^{14}$ | 0.063 | 30 |
| NiFe-HT | 2.29 | 2.10 | 2.56 | 29.11 | 15.1 | 16.8 | 2.0 | 0.76 | 3600 | 1381.5 | 2.01·10$^{40}$ | 0.021 | 24 |

The electrocatalytic OER activity of NiFe-LDH was tested in a basic medium (1 M KOH) in a standard three-electrode cell. For comparative purposes we have prepared working electrodes consisting on NiFe-A and NiFe-HT, as well as commercial IrO$_2$ catalyst as a reference, using powdered samples coated on Ni-foam collectors. Moreover, to overcome the intrinsic insulating behaviour of LDH and in order to improve their catalytic performance and stability, we have directly grown 3D porous films of NiFe-A on Ni-foam –hereinafter: NiFe-A-NiFoam– following a modified procedure previously described in the literature (see SI 6).[13,33] The cyclic voltammetry at different scan rates of NiFe-A

(Figure 7A) reveals the presence of a redox peak around 1.35 V *vs.* RHE, that can be assigned to the Ni(II)/Ni(III or IV) redox processes, probably related with the transformation between $Ni_{1-x}Fe_x(OH)_2$ and $Ni_{1-x}Fe_xOOH$ (see SI7 for additional CVs of the other the samples).[14,33,56] Moreover, the anodic wave of the NiFe-A catalyst is nearly merged with the catalytic wave but a distinct cathodic feature is evident, in excellent accordance with previous reports.[56,57]

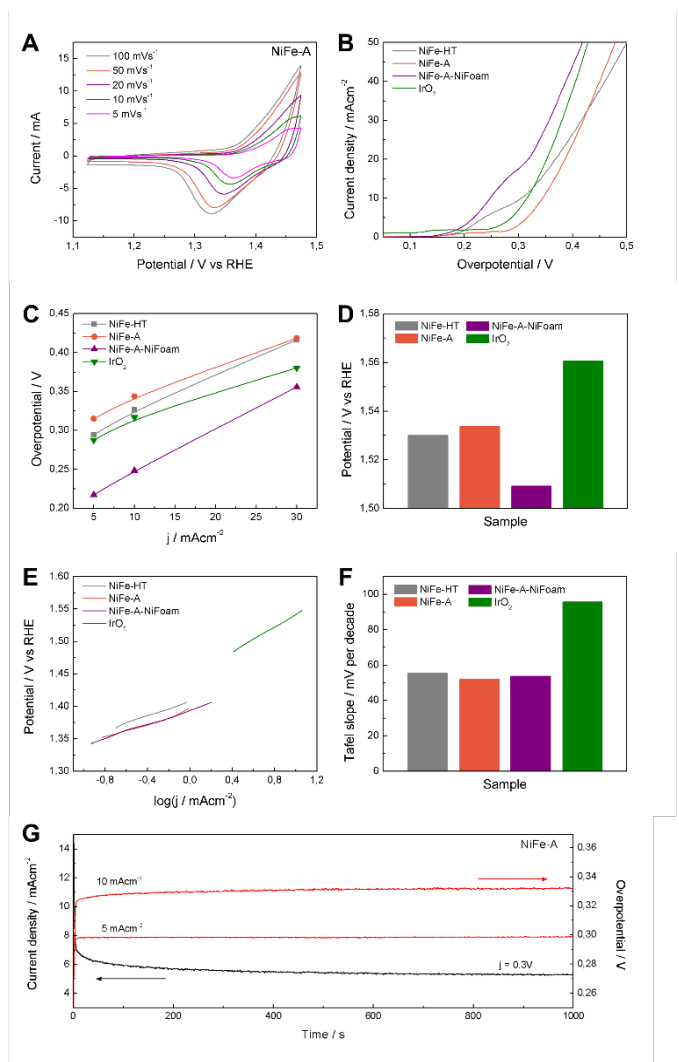

**Fig. 7** (A) Cyclic voltammetry at different scan rates of NiFe-A in 1 M KOH solution. (B) Polarization curves of NiFe-LDHs and commercial $IrO_2$. (C) Overpotential required at different current densities. (D) Onset values of the different NiFe-LDHs and the comercial $IrO_2$ corresponding to the polarization curves. (E) Tafel plots and (F) the histogram of corresponding values of Tafel slopes. (G) The potentiostatic and galvanostatic stability testing under a certain potential or current density.

The catalytic materials were measured by linear sweep voltammetry, showing the lowest onset potential for the NiFe-A-NiFoam (1.51 V *vs.* RHE), followed by the NiFe-HT and the pristine NiFe-A (Fig. 7B and SI7). Different parameters were calculated to quantify the improvements of activity: the overpotential (η) at different current densities (5, 10 and 30 mA·cm$^{-2}$), the current density (j) at η = 300mV and the Tafel slopes (the performance of the different samples have been summarized in SI8). The current density of 10 mA·cm$^{-2}$ was chosen because it represents the current density from a device with 12 % solar to hydrogen efficiency, considered as a realistic measure of the catalytic activity.[15]

As presented in Figure 7C, an overpotential of *ca.* 0.249 V is required at j = 10 mA·cm$^{-2}$ for NiFe-A-NiFoam, a value much smaller than that of NiFe-HT (more than 0.32 V) or NiFe-A (0.34 V) and similar to that previously published by Lu and co-workers.[33] In Figure 7D it can be seen that the overpotential is decreased by 60-80 mV by growing the LDH directly on the Ni-foam. The excellent catalytic activity of the as-synthesized NiFe-LDHs is also reflected in the Tafel slopes, showing values in the range of 52–55 mV per decade, much smaller that that exhibited by the commercial IrO$_2$.

To further evaluate their electrocatalytic activity, the values of the turnover frequency (TOF) of the powdered samples were calculated by assuming that all the transition metal ions in the catalysts are contributing to the reaction, which also confirm that NiFe-A has the highest TOF of 0.01 s$^{-1}$ at an overpotential of 0.3 V, nearly of NiFe-HT (0.007). It is worth to keep in mind that these TOF values compete favourably with those recently reported for others NiFe-LDHs.[26,33]

The stability and durability of the NiFe-A powdered catalyst was tested at constant current densities j of 5 and 10 mA·cm$^{-2}$ and with a constant overpotential of η 300 mV for 1000 s. In Figure 7G, we can see a very high stability in both cases. When increasing the current density from 5 to 10 mA·cm$^{-2}$ the overpotential correspondingly increases to *ca.* 0.03 V (see SI7 for additional measurements). These results are in good accordance with NiFe-LDH nanosheets synthesized using the topochemical approach having the same composition (Ni:Fe 2:1).[26] When compared to Ni(OH)$_2$, the more complex local environment around Ni–O pairs in NiFe-LDHs plays a crucial role in the OER activity, increasing the NiOOH conductivity >30-fold. In fact, Fe exerts a partial-charge transfer activation effect on the Ni atoms, improving the catalytic activity.[56] Moreover, operando Mössbauer spectroscopic studies demonstrated the exclusive formation of Fe$^{4+}$ in NiFe, but not in Fe-only catalysts, as a result of the stabilizing effect of the local NiOOH lattice, most probably on the edges, corners or defects.[57] In any case, the role of the particle size and defect sites in the OER behaviour of NiFe-LDH catalysts remains an open question. These results illustrate the great potential exhibited by NiFe-A as a cheap electrocatalyst for the OER, and opens the door for its hybridization with organic counterparts like graphene or carbon nanotubes in order to further improve its electrochemical performance.

## Conclusions

In summary, we have demonstrated the synthesis of alkoxide-intercalated NiFe-layered double hydroxide exhibiting lateral dimensions of *ca*. 110 nm with a high crystalline quality as revealed by STEM analysis. Moreover, we studied *via* TG-MS measurements its interlamellar composition, proving the methoxide intercalation. Their successful exfoliation in water and 1-BuOH/formamide mixtures has also been performed showing the easy formation of homogeneous ultrathin films with thicknesses in the range of a few nanometers. The magnetic measurements revealed the role exerted by the nanometric dimensions of the platelets, exhibiting superparamagnetic size effects and spin-glass like behaviour with $T_M$ temperatures of *ca*. 8 K. Finally, we have investigated their possible application as water oxidation electrocatalysts. They exhibit an outstanding performance similar to those recently reported for other NiFe-LDHs. This work illustrates the great potential of these earth-abundant 2D materials not only in magnetism or as building blocks for the development of hybrid materials, but also as excellent candidates in the field of energy storage and conversion.

## Acknowledgements

Financial support from the EU (FET-OPEN 2D-INK, grant agreement 664878, Graphene Flagship, grant agreement 604391 and ERC Starting Investigator Award STEMOX, grant agreement 239739), the Spanish MINECO (Project MAT2014-56143-R and Excellence Unit Maria de Maeztu, MDM-2015-0538) and the Generalitat Valenciana (Prometeo and ISIC-Nano Programs) is gratefully acknowledged. Co-funding from UE is also acknowledged. Electron microscopy observations carried out at the Centro Nacional de Microscopía Electrónica, Universidad Complutense de Madrid. We thank the Universidad de Valencia for support from VLC/CAMPUS and INNCIDE program, and for a predoctoral grant (to J. A. C.). G. A. thanks the EU for a Marie Curie Fellowship (FP7/2013-IEF-627386).

## References

1　A. K. Geim and K. S. Novoselov, *Nat. Mater.*, 2007, **6**, 183–191.
2　A. K. Geim and I. V. Grigorieva, *Nature*, 2013, **499**, 419–425.
3　Q. H. Wang, K. Kalantar-Zadeh, A. Kis, J. N. Coleman and M. S. Strano, *Nat. Nanotechnol.*, 2012, **7**, 699–712.
4　V. Nicolosi, M. Chhowalla, M. G. Kanatzidis, M. S. Strano and J. N. Coleman, *Science*, 2013, **340**, 1226419.
5　D. Hanlon, C. Backes, E. Doherty, C. S. Cucinotta, N. C. Berner, C. Boland, K. Lee, A. Harvey, P. Lynch, Z. Gholamvand, S. Zhang, K. Wang, G. Moynihan, A. Pokle, Q. M. Ramasse, N. McEvoy, W. J. Blau, J. Wang, G. Abellan, F. Hauke, A. Hirsch, S. Sanvito, D. D. O'Regan, G. S. Duesberg, V. Nicolosi and J. N. Coleman, *Nat. Commun.*, 2015, **6**, 8563.
6　S. M. Auerbach, K. A. Carrado and P. K. Dutta, *Handbook of Layered Materials*, CRC Press, 2004.
7　V. Rives, Ed., *Layered double hydroxides: present and future*, Nova Science Publishers, Huntington, N.Y, 2001.

8 F. Li and X. Duan, in *Layered Double Hydroxides*, eds. X. Duan and D. G. Evans, Springer Berlin Heidelberg, 2006, pp. 193–223.
9 V. Rives, M. del Arco and C. Martín, *Appl. Clay Sci.*, 2014, **88–89**, 239–269.
10 G. Abellán, C. Martí-Gastaldo, A. Ribera and E. Coronado, *Acc. Chem. Res.*, 2015, **48**, 1601–1611.
11 Q. Wang and D. O'Hare, *Chem. Rev.*, 2012, **112**, 4124–4155.
12 G. Chen, Y. Zhao, G. Fu, P. N. Duchesne, L. Gu, Y. Zheng, X. Weng, M. Chen, P. Zhang, C.-W. Pao, J.-F. Lee and N. Zheng, *Science*, 2014, **344**, 495–499.
13 J. Luo, J.-H. Im, M. T. Mayer, M. Schreier, M. K. Nazeeruddin, N.-G. Park, S. D. Tilley, H. J. Fan and M. Grätzel, *Science*, 2014, **345**, 1593–1596.
14 M. Gong, Y. Li, H. Wang, Y. Liang, J. Z. Wu, J. Zhou, J. Wang, T. Regier, F. Wei and H. Dai, *J. Am. Chem. Soc.*, 2013, **135**, 8452–8455.
15 F. Song and X. Hu, *Nat. Commun.*, 2014, **5**, 4477.
16 H. Wang, Y. Liang, M. Gong, Y. Li, W. Chang, T. Mefford, J. Zhou, J. Wang, T. Regier, F. Wei and H. Dai, *Nat. Commun.*, 2012, **3**, 917.
17 Y. Li, M. Gong, Y. Liang, J. Feng, J.-E. Kim, H. Wang, G. Hong, B. Zhang and H. Dai, *Nat. Commun.*, 2013, **4**, 1805.
18 G. Abellán, E. Coronado, C. Martí-Gastaldo, A. Ribera and J. F. Sánchez-Royo, *Chem. Sci.*, 2012, **3**, 1481–1485.
19 G. Abellán, E. Coronado, C. Martí-Gastaldo, A. Ribera and T. F. Otero, *Part. Part. Syst. Charact.*, 2013, **30**, 853–863.
20 G. Abellán, J. A. Carrasco, E. Coronado, J. P. Prieto-Ruiz and H. Prima-García, *Adv. Mater. Interfaces*, 2014, **1**, 1400184.
21 J. A. Carrasco, H. Prima-Garcia, J. Romero, J. Hernández-Saz, S. I. Molina, G. Abellán and E. Coronado, *J. Mater. Chem. C*, 2016, **4**, 440–448.
22 Y. Han, Z.-H. Liu, Z. Yang, Z. Wang, X. Tang, T. Wang, L. Fan and K. Ooi, *Chem. Mater.*, 2008, **20**, 360–363.
23 G. Abellán, E. Coronado, C. Martí-Gastaldo, E. Pinilla-Cienfuegos and A. Ribera, *J. Mater. Chem.*, 2010, **20**, 7451–7455.
24 G. Abellán, E. Coronado, C. Martí-Gastaldo, J. Waerenborgh and A. Ribera, *Inorg. Chem.*, 2013, **52**, 10147–10157.
25 J.-H. Lee, D. O'Hare and D.-Y. Jung, *Bull. Korean Chem. Soc.*, 2012, **33**, 725–727.
26 W. Ma, R. Ma, C. Wang, J. Liang, X. Liu, K. Zhou and T. Sasaki, *ACS Nano*, 2015, **9**, 1977–1984.
27 G. Abellán, J. A. Carrasco and E. Coronado, *Inorg. Chem.*, 2013, **52**, 7828–7830.
28 E. Gardner, K. M. Huntoon and T. J. Pinnavaia, *Adv. Mater.*, 2001, **13**, 1263.
29 J. A. Gursky, S. D. Blough, C. Luna, C. Gomez, A. N. Luevano and E. A. Gardner, *J. Am. Chem. Soc.*, 2006, **128**, 8376–8377.
30 G. Abellán, J. A. Carrasco, E. Coronado, J. Romero and M. Varela, *J. Mater. Chem. C*, 2014, **2**, 3723–3731.
31 J. H. Lee, S. W. Rhee and D.-Y. Jung, *Chem. Mater.*, 2004, **16**, 3774–3779.
32 I. Horcas, R. Fernández, J. M. Gómez-Rodríguez, J. Colchero, J. Gómez-Herrero and A. M. Baro, *Rev. Sci. Instrum.*, 2007, **78**, 013705–8.
33 Z. Lu, W. Xu, W. Zhu, Q. Yang, X. Lei, J. Liu, Y. Li, X. Sun and X. Duan, *Chem. Commun.*, 2014, **50**, 6479.
34 G. Boehm and M. Dwyer, *J. Chem. Educ.*, 1981, **58**, 809.
35 G. Abellán, P. Amo-Ochoa, J. L. G. Fierro, A. Ribera, E. Coronado and F. Zamora, *Polymers*, 2015, **8**, 5.
36 E. Conterosito, L. Palin, D. Antonioli, D. Viterbo, E. Mugnaioli, U. Kolb, L. Perioli, M. Milanesio and V. Gianotti, *Chem. – Eur. J.*, 2015, **21**, 14975–14986.
37 X. Guo, F. Zhang, D. G. Evans and X. Duan, *Chem. Commun.*, 2010, **46**, 5197.
38 S. Bhattacharya, A. Datta, J. M. Berg and S. Gangopadhyay, *J. Microelectromechanical Syst.*, 2005, **14**, 590–597.
39 K. S. Kim, K. H. Lee, K. Cho and C. E. Park, *J. Membr. Sci.*, 2002, **199**, 135–145.
40 H. W. Nesbitt, D. Legrand and G. M. Bancroft, *Phys. Chem. Miner.*, 2000, **27**, 357–366.
41 P. Rabu, E. Delahaye and G. Rogez, *Nanotechnol. Rev.*, 2015, **4**, 557–580.
42 J. J. Almansa, E. Coronado, C. Martí-Gastaldo and A. Ribera, *Eur. J. Inorg. Chem.*, 2008, **2008**, 5642–5648.
43 J. A. Mydosh, *Spin glasses: an experimental introduction*, Taylor & Francis, London ; Washington, DC, 1993.
44 C. J. Wang, Y. A. Wu, R. M. J. Jacobs, J. H. Warner, G. R. Williams and D. O'Hare, *Chem. Mater.*, 2011, **23**, 171–180.

45   J. Pérez-Ramírez, A. Ribera, F. Kapteijn, E. Coronado and C. J. Gómez-García, *J. Mater. Chem.*, 2002, **12**, 2370–2375.
46   G. Abellán, E. Coronado, C. J. Gómez-García, C. Martí-Gastaldo and A. Ribera, *Polyhedron*, 2013, **52**, 216–221.
47   G. Layrac, D. Tichit, J. Larionova, Y. Guari and C. Guérin, *J. Phys. Chem. C*, 2011, **115**, 3263–3271.
48   G. Abellán Sáez, Hybrid magnetic materials based on layered double hydroxides: from the chemistry towards the applications, PhD Thesis, 2014.
49   Y. Liang, Y. Li, H. Wang and H. Dai, *J. Am. Chem. Soc.*, 2013, **135**, 2013–2036.
50   M. Gong and H. Dai, *Nano Res.*, 2014, 1–17.
51   J. R. Galán-Mascarós, *ChemElectroChem*, 2015, **2**, 37–50.
52   X. Long, J. Li, S. Xiao, K. Yan, Z. Wang, H. Chen and S. Yang, *Angew. Chem. Int. Ed.*, 2014, **126**, 7714–7718.
53   R. Chen, G. Sun, C. Yang, L. Zhang, J. Miao, H. B. Tao, H. Yang, J. Chen, P. Chen and B. Liu, *Nanoscale Horiz.*, 2015.
54   C. Tang, H.-S. Wang, H.-F. Wang, Q. Zhang, G.-L. Tian, J.-Q. Nie and F. Wei, *Adv. Mater.*, 2015, **27**, 4516–4522.
55   Y. Li, H. He, W. Fu, C. Mu, X.-Z. Tang, Z. Liu, D. Chi and X. Hu, *Chem Commun*, 2016. DOI: 10.1039/C5CC08150E.
56   L. Trotochaud, S. L. Young, J. K. Ranney and S. W. Boettcher, *J. Am. Chem. Soc.*, 2014, **136**, 6744–6753.
57   J. Y. C. Chen, L. Dang, H. Liang, W. Bi, J. B. Gerken, S. Jin, E. E. Alp and S. S. Stahl, *J. Am. Chem. Soc.*, 2015, **137**, 15090–15093.

# Supplementary Information

**Alkoxide-intercalated NiFe-layered double hydroxides magnetic nanosheets as efficient water oxidation electrocatalyst**

Jose A. Carrasco,[a] Jorge Romero,[a] María Varela,[b] Frank Hauke,[c] Gonzalo Abellán,[a,c,*], Andreas Hirsch[c] and Eugenio Coronado[a,*]

[a] Instituto de Ciencia Molecular (ICMol), Universidad de Valencia, Catedrático José Beltrán 2, 46980, Paterna, Valencia, Spain

[b] Universidad Complutense de Madrid, Dpt. Fisica Aplicada III & Instituto Pluridisciplinar, Madrid 28040, Spain

[c] Department of Chemistry and Pharmacy and Institute of Advanced Materials and Processes (ZMP), University Erlangen-Nürnberg, Henkestr. 42, 91054 Erlangen and Dr.-Mack Str. 81, 90762 Fürth, Germany.

**Contents**

**SI 1.** FESEM image of the NiFe-A sample.

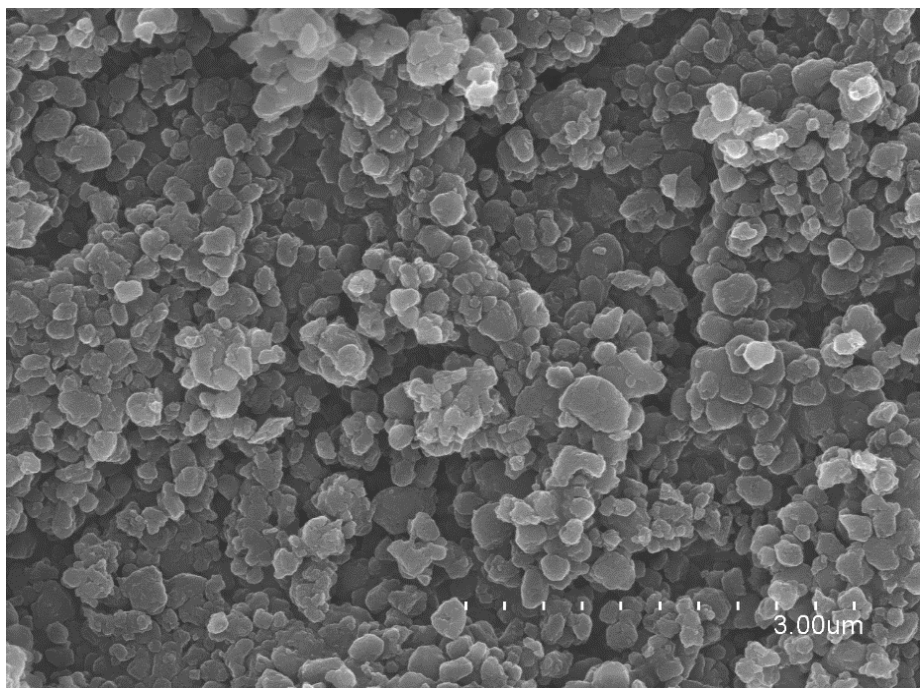

**SI 2.** FESEM image of the NiFe-HT sample.

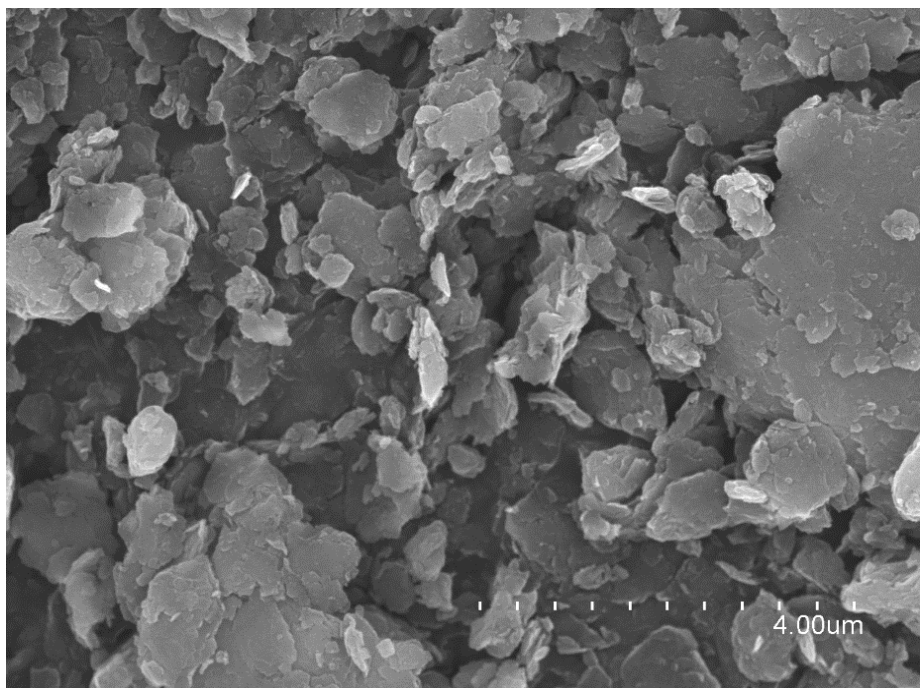

**SI 3.** AFM images highlighting the coverage of the NiFe-A sample.

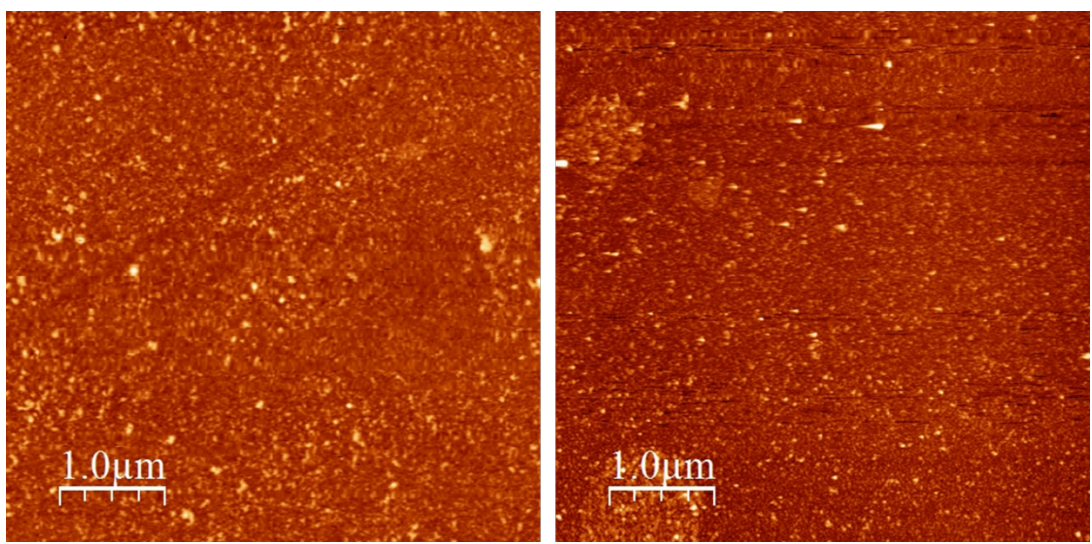

As explained in the main text, the deposition procedure has been carried out with a mixture of formamide/BuOH and with a $SiO_2$-Si substrate previously activated with $O_2$ plasma.

**SI 4.** AFM characterization of a SiO$_2$-Si substrate (blank).

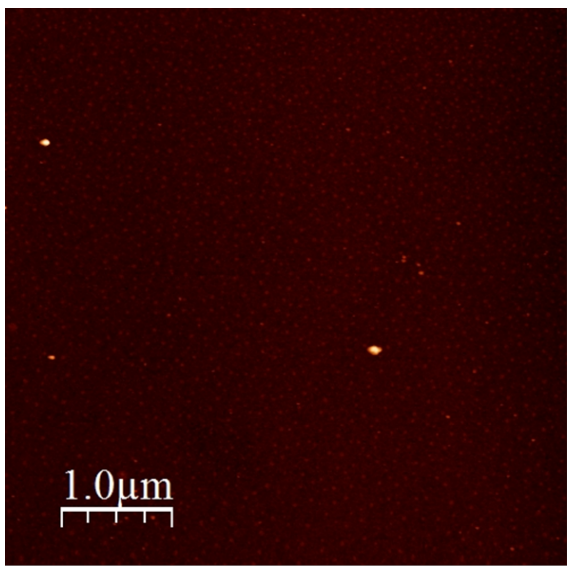 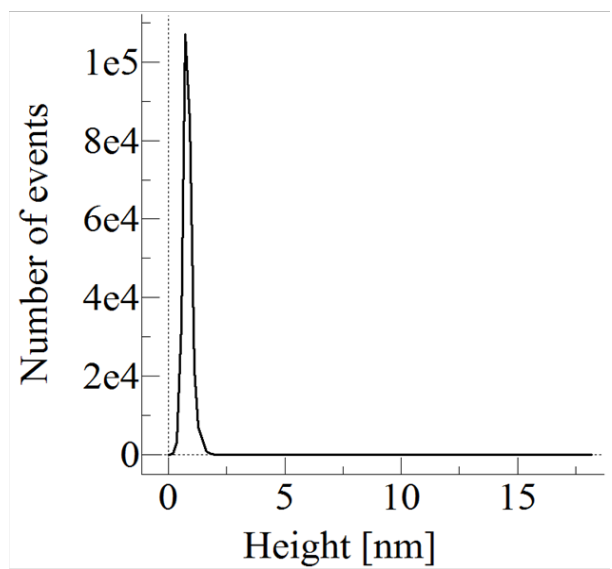

RMS value = 0.15 nm.

Average height = 0.90 nm

**SI 5.** Arrhenius plot and 3D scaling law model of the spin dynamics for the NiFe-HT sample.

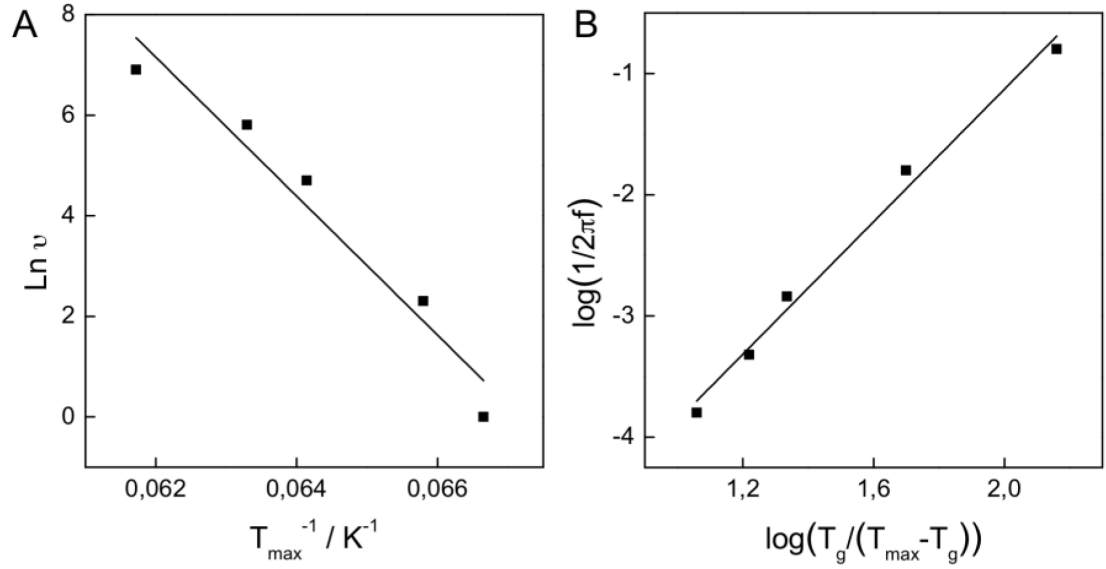

(A) Arrhenius fitting of the $\chi''_M$ signal for the NiFe-HT sample. (B) Frequency dependence of $\chi''_M$ fitted with the 3D scaling low model $\tau = \tau_0 \cdot [T_g/(T_g/T_{max} - T_g)]^{z\nu}$.

**SI 6.** SEM images of the NiFe-A-NiFoam sample.

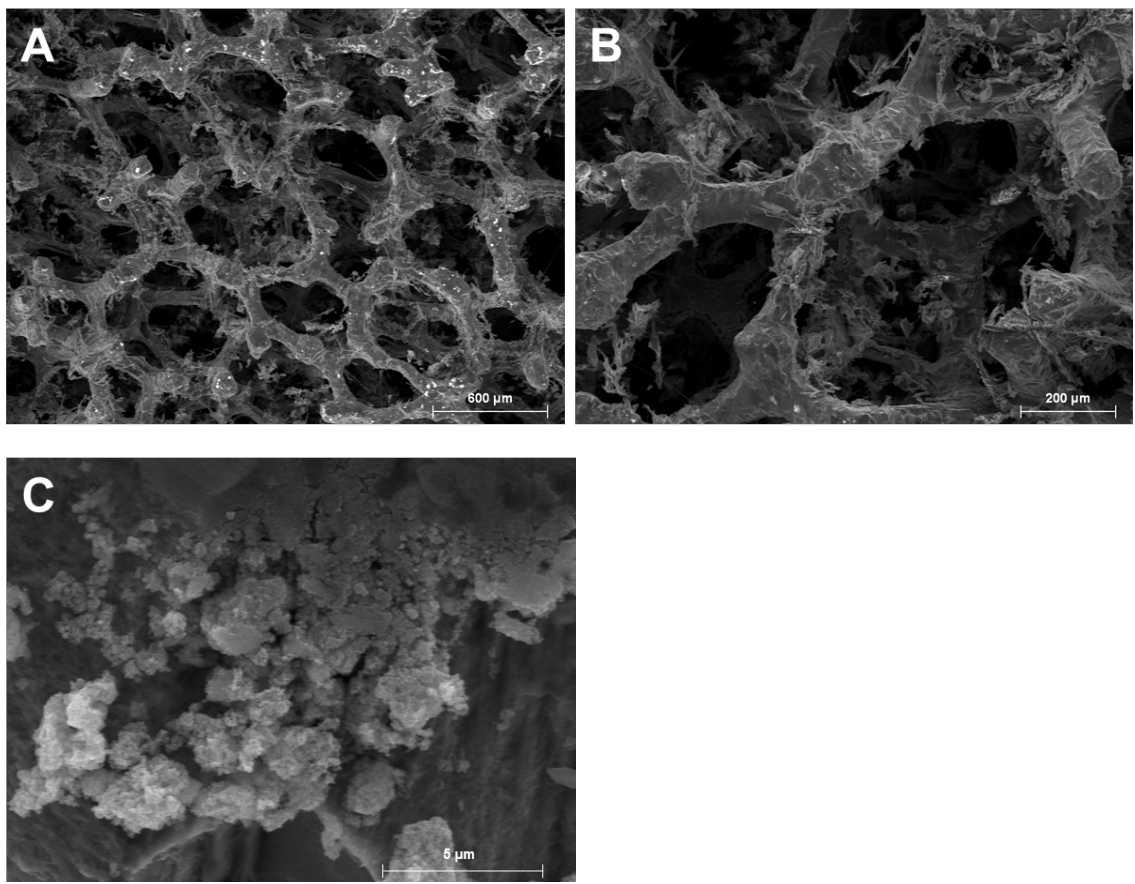

**SI 7.** Additional electrochemical measurements.

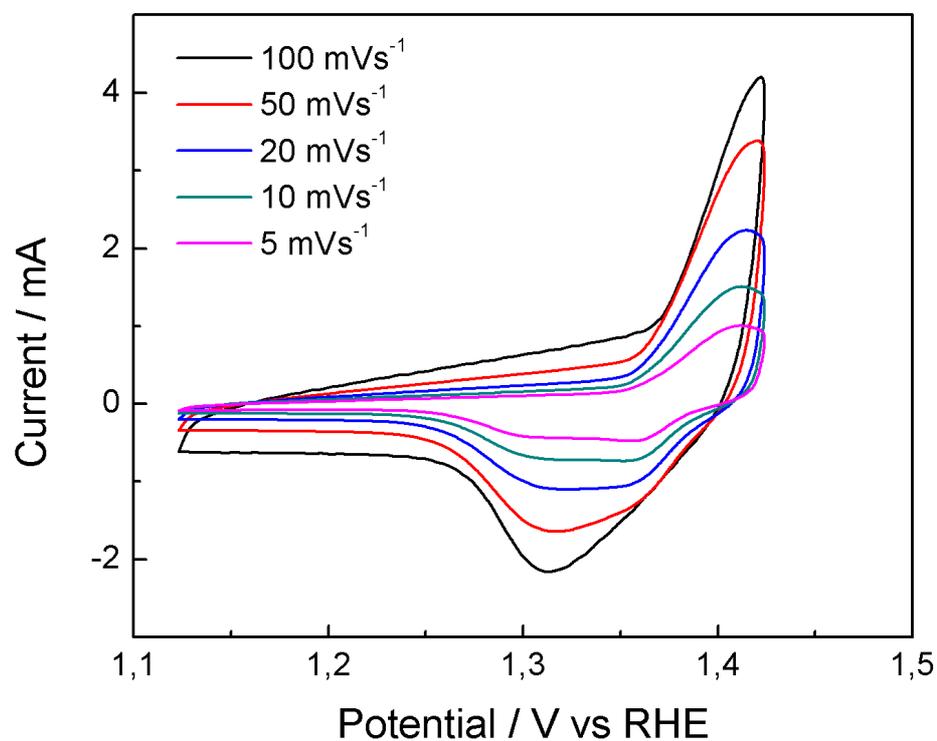

**Figure SI 7.1.** CV of NiFe-A-NiFoam at different scan rates.

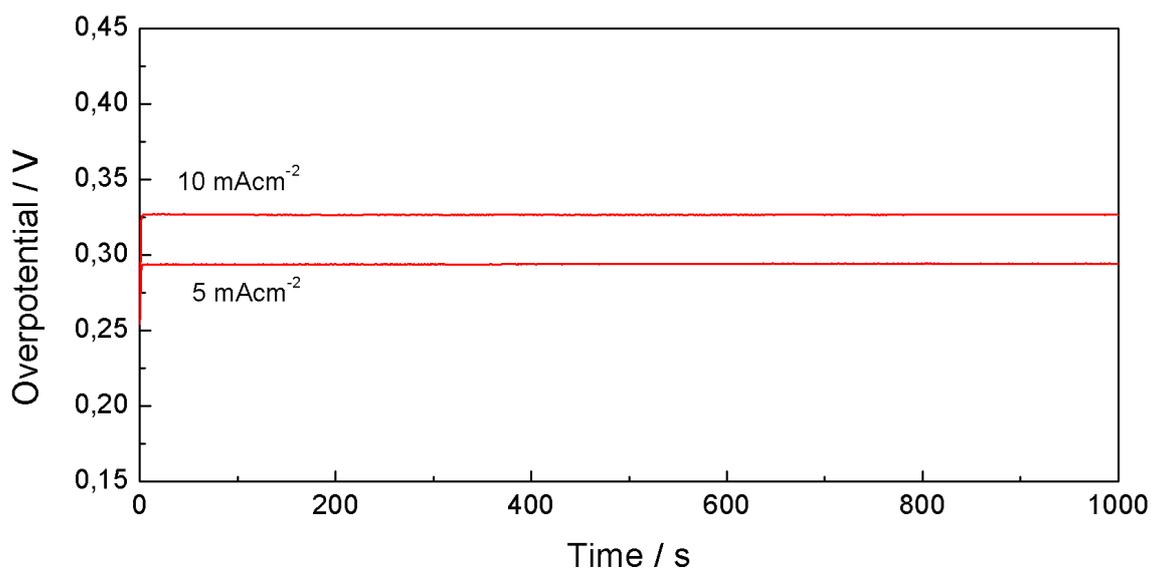

**Figure SI 7.2.** Potentiostatic stability testing of NiFe-A-NiFoam under a certain current density.

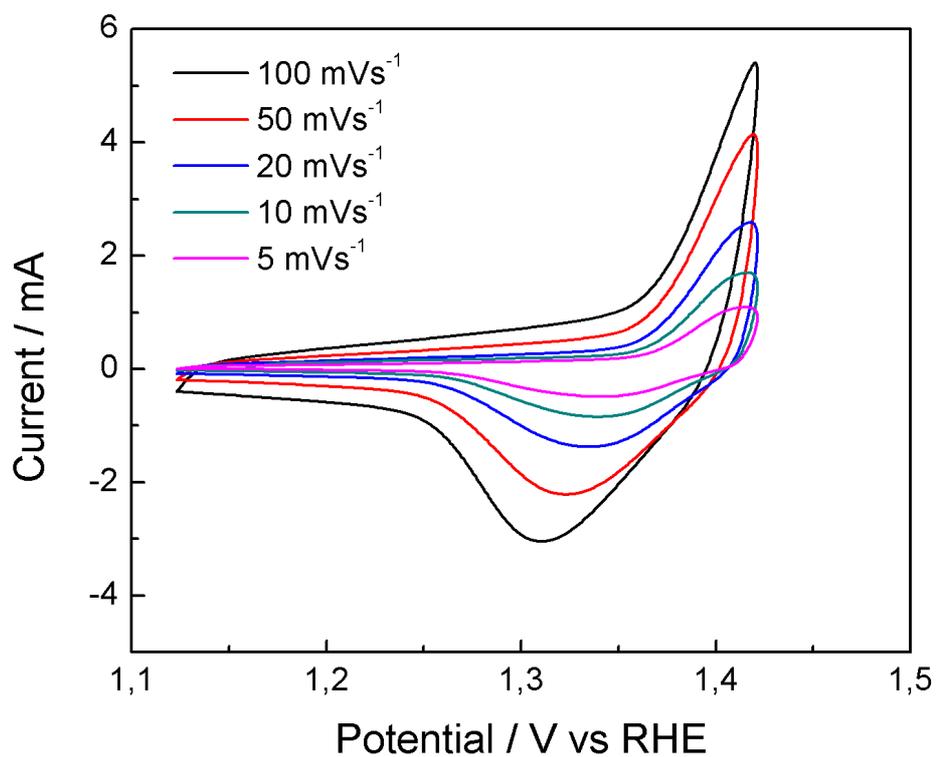

**Figure SI 7.3.** CV of NiFe-HT at different scan rates.

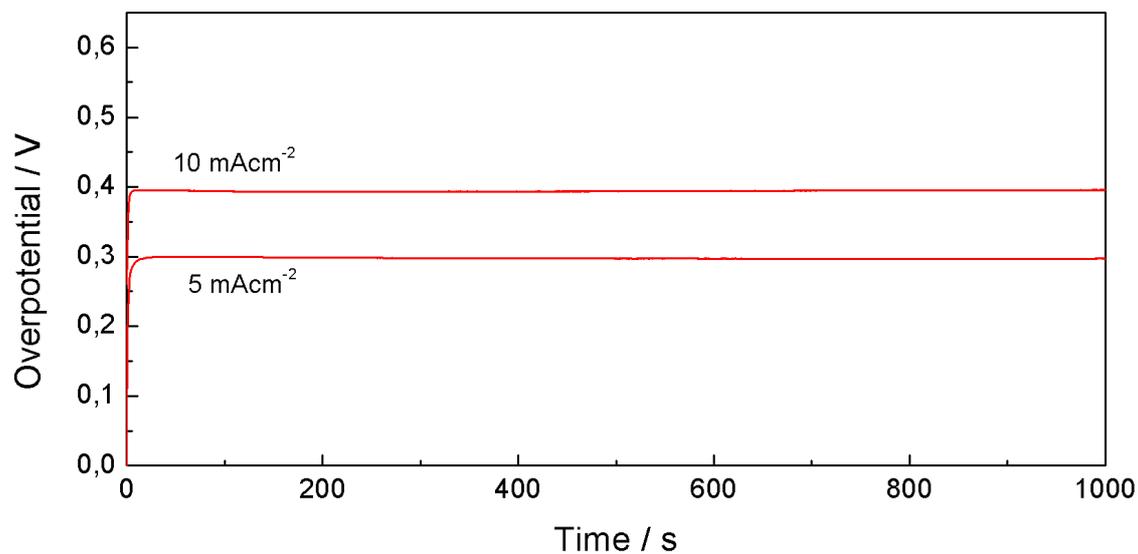

**Figure SI 7.4.** CV Potentiostatic stability testing of NiFe-HT under a certain current density.